\documentclass[9pt]{article}

\begin{document}
\date{}

\def\a{\alpha}
\def\la{\lambda}
\def\si{\sigma}
\def\z{\zeta}
\def\phi{\varphi}
\def\ka{\kappa}

\def\pd{\partial}
\def\d{{\rm d}}       
\def\~#1{\widetilde #1}
\def\.#1{\dot #1}
\def\^#1{\widehat #1}
\def \q{\quad}
\def\={\, =\, }

\def\beq{\begin{equation}}
\def\eeq{\end{equation}}

\def \sy {symmetry}
\def \sys {symmetries}
\def \eq {equation}
\def\ni{\noindent}
\newtheorem{theorem}{Theorem}
\newtheorem{Proposition}{Proposition}


\title{On the connections between symmetries and conservation rules
of dynamical systems}

\author{
   Giampaolo Cicogna\thanks{Email: cicogna@df.unipi.it} 
   \\~\\
Dipartimento di Fisica ``E.Fermi'' dell'Universit\`a di Pisa\\
  and  Istituto Nazionale di Fisica Nucleare, Sez. di Pisa \\~\\
Largo B. Pontecorvo 3, Ed. B-C, I-56127, Pisa, Italy  }

\maketitle

\begin{abstract}
The strict connection between Lie point-symmetries of a 
dynamical system and its constants of motion is discussed and emphasized, through
old and new results. It is shown in particular how the knowledge of a 
symmetry of a dynamical system can allow to obtain 
conserved quantities which are invariant under the symmetry. 
In the case of Hamiltonian dynamical systems it is shown that, if the
system admits a symmetry of ``weaker'' 
type (specifically, a $\lambda$ or a $\Lambda$-symmetry), then the generating function 
of the symmetry is not a conserved quantity, but the deviation from the exact 
conservation is ``controlled'' in a well defined way.
Several examples illustrate the various aspects.
\end{abstract}

\bigskip \ni

\medskip
\ni
{\it Keywords}: {Dynamical Systems; Lie point-symmetries; 
$\lambda$-symmetries; \\ constants of motion; Hamiltonian dynamical systems;
conservation rules}

\bigskip\ni
{\bf Talk given at the ICNAAM Conference, Halkidiki (Greece), \\ September 2011}

\smallskip

\section{Introduction}

The role and the relevance of methods based on the analysis of symmetry
properties of differentialequations (both ordinary and partial) are well 
known, for what concerns not only the problem of finding explicit solutions, but 
also of examining ``structural'' properties (a typical and relevant 
feature is, e.g., the presence of conservation rules).

There is an enormous literature on this subject: see e.g. 
\cite{Ovs,Olv,CRC,BA} for some classical texts where general procedures and 
standard applications can be found. However, there is a particular 
context where \sy\ methods meet some intrinsic difficulty: this is the 
case of dynamical systems (DS), i.e.  systems of first-order 
time-evolution differential \eq s of the form
\beq\label{DS1}\.u_a\=f_a(u,t)\q\q\ u_a=u_a(t)\q\ (a=1,\ldots, n) \eeq
with $\.u=\d u/\d t$ and (sufficiently smooth) given functions 
$f_a=f_a(u,t)$. 
The present paper is devoted to investigate precisely this case. 
I am referring more specifically to Lie point-\sys 
, i.e. to continuous transformations generated by infinitesimal vector fields $X$ 
which can be written in the form
\beq\label{Xt} X\=\phi_a(u,t)\frac{\pd}{\pd u_a}+\tau(u,t)\frac{\pd}{\pd t}
\equiv \phi\cdot\nabla_u+\tau\pd_t\ .\eeq
The problem of finding all  \sys\ admitted by a DS is 
quite difficult. On the other hand, expectedly, the presence of some \sy\ 
is strictly related to the determination of first integrals of the 
differential problem, also called, in this context, constants of motion of the 
dynamical flow. The determination of constants of motion for a time-evolution process 
is clearly a basic result, not only in view of obtaining full solutions of 
the problem, but also for their physical interpretation as ``conserved 
quantities'' along the time evolution.

Sect.2 is devoted to present the \eq s providing the conditions for the 
existence of Lie point-\sys\ of a given DS, and to discuss the close 
connection existing between the problem of solving these \eq s and of 
detecting constants of motion. It can be noticed that this connection, in the 
present context, 
is not {\it directly} related to the celebrated Noether theorem. It will be 
useful, instead, in view of our discussion, to recall two classical, perhaps 
less known, 
old results.  In few words, these results are very interesting as they 
illustrate the strict relationship existing, also in the context of DS, 
between the 
notions of \sys\ and of constants of motion, but on the other hand they are not very 
useful in practice because they require the knowledge of ``many'' constants of motion in order 
to find (possibly all) \sys\ of the DS. 

In Sect.3, the point of view is partly 
reversed: I will assume that {\it just only one} \sy\ is known.  
Indeed, although finding all \sys\ can be a difficult problem, 
it often happens that one is able to detect a single \sy , and 
often this \sy\ has a rather simple expression. Sect.4 is devoted to show 
two quite general new classes of examples where this situation occurs. The 
first one includes Lorenz-like DS, the second one deals with systems 
related to higher order ODE's.

I will then show that this \sy\ can be used to introduce some suitable 
``\sy -adapted coordinates'' where the problem of determining the 
constants of motion becomes easier and provides conserved quantities which are also invariant under 
the \sy . This is completely confirmed by the examples 
considered in Sect.4.

In addition, it is shown that a ``weaker'' notion of \sy\ can be 
introduced to the same purpose. I am referring to the notion of 
$\la$-\sy\ \cite{MR1,MR2}, which  will be briefly recalled especially for what 
concerns its application to DS \cite{MRV,PLA}. In particular, the presence 
of a $\la$-\sy\ allows the introduction of suitable coordinates exactly 
as standard \sys .

In Sect.5, the particular case of {\it Hamiltonian} DS will be considered. 
A classical result ensures that if such a DS exhibits a \sy\ admitting a generating 
function $G$, this 
function is automatically a conserved quantity invariant under the \sy . 
If instead the DS admits a $\la$-\sy , the generating function is no 
longer a conserved quantity, but the ``breaking'' of the conservation is 
``controlled'' in a well defined way. In this context, also a 
generalization of the notion of $\la$-\sy\ will be usefully introduced \cite{PLA,JN}. 
Two examples illustrate the results, with the explicit calculation of the 
``deviation'' from the exact conservation rule.

This is a full paper presented within ICNAAM 2011; a very short and preliminary 
sketch of part of these results can be found in the enlarged 
Abstracts of the Conference Proceedings \cite{Proc}.

\section{Symmetries and constants of motion: some classical facts}

Following  the standard procedure, see e.g. \cite{Olv,BA}, a vector field $X$ 
(\ref{Xt}) is a Lie point-\sy\ for the DS (\ref{DS1}) (according to an 
usually accepted abuse of language, I will denote by $X$ both the \sy\ and 
its infinitesimal generator) if the following condition is satisfied
(sum over repeated indices unless otherwise stated)
\beq\label{syt}[\,f, \phi\,]_a\=-\frac{\pd}{\pd 
t}(\phi_a-\tau\,f_a)+\frac{\pd\tau}{\pd 
u_b}f_af_b\q\q\q (a,b=1,\ldots,n)\eeq
where $[\,f, \phi\,]_a$ is defined by
\[ [\,f, \phi\,]_a\= f_b\frac{\pd}{\pd u_b}\phi_a-\phi_b\frac{\pd}{\pd 
u_b}f_a \equiv (f\cdot\nabla)\phi_a-(\phi\cdot\nabla) f_a\ .\]
It is not restrictive to put $\tau=0$, possibly introducing
``evolutionary''  vector field 
\[X_{\rm ev}\! := (\phi-\tau\.u)\cdot\nabla\=(\phi-\tau 
f)\cdot\nabla\,\equiv\~\phi\cdot\nabla\]
so the \sy\ condition becomes
\beq\label{sy}[\,f,\phi\,]\,_a+\frac{\pd}{\pd t}\phi_a\=0\ .\eeq
Despite this apparently simple form, it is in general very difficult to 
obtain a complete solution to this set of determining \eq s.
{\it In principle } there are $n$ (functionally independent, locally 
defined) solutions $\phi^{(a)}$; denoting by $\ka=\ka(u,t)$ any  
constant of motion  of the DS, i.e. any function such that
\[D_t\ka\equiv
\pd_t\ka+f\cdot\nabla\ka\=0 \]
the most general \sy\ of the DS can be written as
\[X\=\sum_{a=1}^n\ka^{(a)}\phi^{(a)}\cdot\nabla\equiv \ka^{(a)}X^{(a)}\ .\]
Apart from this very general result, the relationship between 
\sys\ and constants of motion is actually much closer. To illustrate this point, and 
also in view of our discussion, let me recall the 
two following, perhaps less known, classical results.

\bigskip\noindent
{\bf a}. This result is due to Ovsjannikov \cite{OvNov} and shows 
how \sys\ of a DS can be deduced from the knowledge of its constants of motion.

\begin{Proposition} Assume that $n$ functionally independent constants of motion $\ka^{(a)}$ 
of the given DS are known; then the linear system of $n^2$ \eq s 
\[\sum_{a=1}^n p_{ab}\frac{\pd \ka^{(a)}}{\pd u_c}\=\delta_{bc}\]
can be solved for the $n^2$  quantities $p_{ab}$. Then
\[X^{(a)}\=\sum_{b=1}^n p_{ab}\frac{\pd}{\pd u_b}\]
are $n$ independent   \sys\ for the DS.
\end{Proposition}

\bigskip\noindent
{\bf b}. The following result, based on the notion of Liouville vector field, has been 
restated by G. \"Unal \cite{Un}, and also used by J. Zhang  and  Y. Li \cite{ZL}. 
Let me recall the main fact in the following form.

A DS is said to admit a {\it Liouville vector field} $Y=\psi(u,t)\cdot\nabla$ 
if
\[\pd_t\psi_a+[f,\psi]_a+({\tt Div}\,f)\psi_a\=0\]
Clearly, if ${\tt Div}\,f=0$ then $Y$ is a standard \sy ; 
if instead ${\tt Div}\,f\not=0$, putting 
$$Y=q\,X$$ 
then $X$ is a standard \sy\ for the DS if $q$ is a scalar function solving
\[\pd_tq+f\cdot\nabla q+({\tt Div}\,f)q=0\ .\]

Then one has:

\begin{Proposition}
If the Liouville vector field $Y$ satisfies
${\tt Div}\,\psi=0$, then there are $n-1$ constants of motion $\^\ka^{(a)}$ such that 
\beq\label{ep} Y\= \psi_a\frac{\pd}{\pd 
u_a}\=\varepsilon_{abc\ldots
l}\^\ka^{(1)}_{,b}\,\^\ka^{(2)}_{,c}\ldots\^\ka^{(n-1)}_{,l}\frac {\pd}{\pd
u_a}
\eeq
where $\^\ka^{(a)}_{,b}=\pd\^\ka^{(a)}/\pd u_b$.
In addition,
the above constants of motion  $\^\ka^{(a)}$ are invariant
under both the vector fields $Y$ and $X$: 
\[Y\^\ka^{(a)}=X\^\ka^{(a)}\=0\ .\]
\end{Proposition}
\medskip
The last sentence says that the  quantities $\^\ka^{(a)}$ are 
simultaneously 
invariant under the dynamical flow and under the \sy  
\footnote{
In \cite{Un} this result is stated saying that the quantities 
$\^\ka^{(a)}$ appearing in (\ref{ep}) are `the' first integrals of the DS. 
Clearly, not all the first integrals satisfy (\ref{ep}) nor are \sy 
-invariant.}.

\bigskip
Both the above results are conceptually greatly relevant, but clearly 
of little practical use if one wants to explicitly find \sys\ (or constants of motion
as well) of a given DS. In the following, I try to partly reverse the 
approach: 
I will assume that just {\it only one} \sy\ is known, and then try to 
deduce any possible information from it.

\section{Symmetry adapted coordinates}

Often, given a DS,   one \sy\ of it is easily seen, either by direct 
inspection or by simple calculations, as we shall see in the following 
section. Then (remembering also Proposition 2) the idea is to 
use invariants under this \sy\ to construct (one or more) constants of motion.

To this purpose, the presence of a $\la$-\sy\ (instead of a standard one) 
may equally well help in the 
calculations. Let me briefly recall the basic definitions of $\la$-\sy\ 
for what concerns the application in this context.

The notion of $\la$-\sy\ has been
originally introduced in 2001 by C. Muriel \& J.L.Romero in the context 
of ODE's \cite{MR1,MR2}. Since then, this notion  has received many very 
important applications and extensions, which 
cannot be recalled here (for a fairly complete list of references, 
see e.g. \cite{Gtw,GaC}). In our case, a DS admits a 
$\la$-\sy\ $X=\phi\cdot\nabla$ if there is a $C^\infty$ function 
$\la=\la(u,\.u,t)$ 
such that the following condition holds
\beq\label{ll} [\,f,\phi\,]\,_a+\frac{\pd}{\pd t}\phi_a\=-\la\,\phi_a \eeq
to be compared with the standard condition (\ref{sy}).

$\la$-\sys\ are not properly \sys , indeed, e.g., they do not transform 
solutions into other solutions, 
nevertheless they share with standard \sys\ many useful properties;
in particular they indicate, as well as standard \sys , 
a convenient choice of variables in view of our procedure. 

Let us assume then that the given DS admits either a standard or a 
$\la$-\sy\ $X=\phi\cdot\nabla$. Introduce then $n$ functionally independent 
quantities which are left fixed by this \sy : choose the time $t$ as 
one of these, and the remaining $n-1$, denoted by $w_j=w_j(u)$, independent 
of $t$:
\[ X\,w_j\=X\,t\=0 \q\q (j=1,\ldots,n-1)\ .\]
Let $\z$ be the ``rectifying'' coordinate along the action of $X$, i.e.
\[X\,\z\=1\q\q {\rm or}\q\q X\=\frac{\pd}{\pd \z}\ .\]
Notice that, even in the case of $\la$-\sy , all these coordinates depend 
only on $X$ and not on   $\la$. Choose now $w_j,\z$ as new dependent 
variables (with $t$ still as independent one), and rewrite the DS 
in terms of these, i.e. in the form $\.w_j=W_j,\,\.\z=Z$:  one immediately has that 
the r.h.s. $W_j,Z$ of the new DS turn out to be independent of $\z$ if the 
\sy\ is 
standard \cite{Olv}, and that only $Z$ may depend on $\z$ if the \sy\ is 
a $\la$-\sy\ \cite{MRV,PLA}: this can be summarized writing 
\beq\label{wz} \.w_j  \=W_j(w,t)   \q  , \q \.\z \=Z(w,[\z],t)\ .\eeq
If now we look for constants of motion of the DS {\it expressed as functions of} 
$w_j,\z,t$,
i.e. $\ka=\ka(w,\z,t)$, we can conclude with the following
\begin{Proposition}
The constants of motion $\ka^{(a)}(w,\z,t)$ solve the characteristic \eq
\[\frac{\d w_1}{W_1}\=\ldots\= \frac{\d w_{n-1}}{W_{n-1}}\=\frac{\d\z}{Z}\=\d t\]
where $W_j,Z$ are defined in (\ref{wz}).
\end{Proposition}
The advantage of this procedure is clear: we have a {\it reduction} of 
the initial problem to a system of $n-1$ \eq s involving $n-1$ variables 
$w_j$; for the same reason, also the search for the constants of motion through the above
characteristic \eq\ is easier. One obtains in this way, by 
construction, conserved quantities which are also \sy -invariant;  this 
agrees of course with the classical Frobenious theorem \cite{Olv}.
Notice that this is a special case of a 
more general problem of finding suitable reduction procedures of DS; this 
and other related aspects will be discussed in a paper by G. Gaeta, S. 
Walcher and the present author (in preparation).

\section{Two classes of examples}

The two following propositions provide two quite general classes of DS 
where the presence of one \sy\ is guaranteed and  constants of motion can be successfully 
deduced.

\begin{Proposition} 
A DS of the form
$$  \.u_a\=\si_{(a)}u_a+g_a(u) \q\q (a=1,\ldots,n; \
{\rm no\ sum\ over}\ a)$$
where  $\si_{(a)}$=const  ($\not= 0$) admits the \sy
$$X\=\exp(\la t)\,g\cdot\nabla$$
$\left(\la=const\ (possibly\ zero)\right)$  if $g_a(u)$ have the form
\[ g_a(u)\=u_a^{ \left(1-(\la/\si_{(a)})\right)} P_a(u)
\q\q (a,b,c=1,\ldots,n)\]
where $P_a$ are any smooth functions of the ratios 
$u_b^{\si_{(c)}}/{u_c^{\si_{(b)}}}$ with 
``exchanged'' exponents.
\end{Proposition}

Several examples of this situation are known: it includes e.g. generalized 
Lorenz systems \cite{ZL}, etc.

\medskip\noindent
{\it Remark.} If $\la\not=0$, one can equivalently say that  
$X':=g\cdot\nabla$ is a {\it $\la$-\sy } 
with $\la(u,\.u,t)=\la=$ const. 
This follows from the general property that
if $X_\la$ is a $\la$-\sy\ with some  $\la(u,\.u, t)$ then 
\[X:=\exp\left(\int\la(u,\.u,t)\,\d t\right)X_\la\]
is a (possibly {\it nonlocal}) standard   \sy  
\footnote{According to this remark, all examples given in \cite{ZL}
are actually equivalent to standard (not properly $\la$) \sys .}.

\medskip
Here an explicit example for Proposition 4.

\medskip
\noindent
{\it Example 1. }
With $n=3$ and $u\equiv(x,y,z)$ consider the DS:
\[ \.x=-x+x^2P_1 \  ,\  \.y=-y+y^2P_2 \  ,\ 
\.z=-2z+z^{(3/2)}P_3\]
where $P_a$ are functions of $x/y,x^2/z$, and which can be cast in analytic form
\[\.x\!=\!-x+Q_1(xy,y^2,z)\  ,\   \.y\!=\!-y+Q_2(x^2,xy,z)\  ,\   \.z\!=
\!-2z+Q_3(x^3,yz,xz)\ .\]
This DS admits the \sy
\[X\=e^t\left(Q_1\frac{\pd}{\pd x}+Q_2\frac{\pd}{\pd y}+
Q_3\frac{\pd}{\pd z}\right)\ .\]
Choose e.g. (this is a variant of an example given in \cite{Un}) 
$$Q_1=z-2y^2\q ,\q Q_2=2xy \q ,\q Q_3=4xz$$ 
then, with the notations of the above section,
\[w_1=z/y^2\q , \q w_2\=x^2+y^2-z/2\ ,\]
\[\z\=\frac{e^{-t}}{4v}\log\Big|\frac{x-v}{x+v}\Big| \q\q \q 
({\rm here} \q v=|w_2|^{1/2})\ .\]  
Following the above procedure, the DS becomes
\[\.w_1=0\q , \q\.w_2=-2w_2\q , \q\.\z=e^{-t}\] 
and  exactly {\it three} functionally independent  constants of motion can be found
\[\ka_1=w_1=z/y^2\q , \q \ka_2=e^{2t}(x^2+y^2-z/2)\]
\[\ka_3=e^{-t} \left(1+\frac{1}{4v}\log\Big|\frac{x-v}{x+v}\Big|\right)\ .\]

\bigskip
\begin{Proposition}
Let $n=2,\,u\equiv(x,y)$; the DS
\[\.x\=y\q\q\.y\=y^2\gamma^{-1}\gamma_x+\gamma\,F(\gamma^{-1}y)\]
where $\gamma=\gamma(x)\not=0$ and $F$ are any given smooth 
functions, admits the \sy 
\[X\=\gamma\frac{\pd}{\pd x}+y\gamma_x\frac{\pd}{\pd y}\ .\] 
\end{Proposition}
Here one has
\[w\=\gamma^{-1}y\q\q , \q\q \z\=\int\gamma^{-1}\d x\]
and the DS becomes 
\[\.w\=F(w)\q\q ,\q\q \.\z\=w\ .\]
Two  constants of motion are easily obtained:
\[\ka_1\=\z-\int w\,F^{-1}(w)\,\d w \q\q ,\q\q \ka_2\=t-\int F^{-1}(w)\,\d w\ .\]
This type of DS is specially interesting because the DS is equivalent to 
the ODE 
\[\ddot x\=\.x^2\gamma^{-1}\gamma_x+\gamma\,F(\gamma^{-1}\.x)\]
and the above \sy\ of the DS is in this case automatically extended to 
become a \sy\ for the ODE:
\[ X_{ODE}\=\gamma\frac{\pd}{\pd x}\=\frac{\pd}{\pd\z }\ .\]
Notice that the ODE becomes just $\ddot\z=F(\dot\z)=F(w)$. 
Similarly, constants of motion for the DS become first integrals for the ODE simply 
replacing $y$ with $\.x$.

It is clearly possible to extend in a suitable way this example to DS and to the 
corresponding ODE to the case $n>2$.

An explicit example follows.

\medskip\noindent
{\it Example 2. }
Choosing $\gamma=e^x,\,F=-y^2\,e^{-2x}=-w^2$, the DS is
\[\.x\=y\q\ , \q\ \.y\=y^2\left(1-e^{-x}\right)\]
with \sy\ $X=\exp(x)(\pd_x+y\pd_y)$, and the ODE is
\[\ddot x\=\.x^2-e^{-x}\,\.x^2\ .\]
Thanks to the new variables, the general solution is easily get
and two constants of motion for the DS (and for the ODE, replacing $y$ with $\.x$) are
\[ \ka_1\=\log|y|-x-e^{-x}\q\ , \q\ \ka_2\=t-e^x/y\ .\]

\section{Hamiltonian DS, $\Lambda$-\sys\  and ``controlled failure'' of  
conservation rules}

Let me now consider the specially interesting case in which the DS is a 
{\it Hamiltonian} DS, i.e. the DS is obtained from a given Hamiltonian 
function $H$.
Changing accordingly the notations, with $n=2m$, the $n$ variables 
$u=u_a(t)$ are replaced by the canonical variables $q_\a(t),p_\a(t)  
\,(\a=1,\ldots,m)$:
\[u\equiv(q_1,\ldots,q_n,p_1,\ldots,p_n)\equiv(q,p)\in{\bf R}^{2m}\]
and the DS is  the system of the Hamilton \eq s of motion  for the given Hamiltonian  
$H=H(q,p,t)$: 
\[\.u\=J\nabla H\=F(u,t) \q\ ; \q\ \nabla\equiv(\nabla_q,\nabla_p)\]
where
\[J\=\pmatrix  {0 & I_m \cr  -I_m & 0} \]
is the standard symplectic matrix and $I_m$ the $m\times m$ identity 
matrix. In the same way, vector fields $X$ will be written
\[X\=\phi_\a(q,p,t)\frac{\pd}{\pd q_\a}+\psi_\a(q,p,t)\frac{\pd}{\pd 
p_\a}\equiv\Phi\cdot\nabla  \q\q ;\q\q \Phi\equiv(\phi_\a,\psi_\a).\]

I will now restrict the attention  on vector fields admitting a {\it generating
function} $G=G(q,p,t)$, i.e. vector fields $X$ satisfying
\[\Phi\=J\nabla G \q {\rm or} \q \phi=\nabla_p G,\,\psi=-\nabla_q G\ . \]
The connection between \sys\ and constants of motion in the Hamiltonian context is even 
more stringent. It is well known indeed, since the end of XIX century \cite{LCiv} 
(see also, e.g., \cite{Olv}), that if $X$ is a Lie 
point-\sy\ for a Hamiltonian DS, i.e. $[F,\Phi]+\pd_t\Phi=0$, then
\beq\label{DTG} \nabla (D_tG)\=0 \q\q {\rm or}\q\q D_tG\=g(t)\ .\eeq
This follows from the identity
\beq\label{LP} \nabla_a (D_tG)=\nabla_a(\{G,H\}+\pd_tG)=-J_{ab}([F,\Phi ]_b
+\pd_t\Phi_b) \q\q (a,b=1,\ldots,n)\eeq
where $\{\cdot,\cdot\}$ is the standard Poisson bracket. 

Then $G$ is a constant  of motion apart from an additional time dependent function  $g$. 
Let me now consider for simplicity  generating functions $G(q,p)$ not depending 
explicitly on $t$; combining this with the obvious property $X(G)=~0$ if 
$X=J\nabla G\cdot\nabla$, the following standard result can be stated for 
convenience
\begin{Proposition}
Let the Hamiltonian DS $\.u=J\nabla H$ admit a \sy\ $X=\Phi\cdot\nabla$ where 
$\Phi=J\nabla G$. Then the generating function $G(q,p)$ is a conserved 
quantity invariant under $X$.
\end{Proposition}
In terms of our previous arguments, $G$ can then be chosen as one of \sy 
-invariant variables $w$, and in this case it is automatically (and 
trivially) also a constant  of motion.

A less trivial and   more interesting situation occurs if the 
Hamiltonian DS does admit some $X$ as a $\la$-\sy : in this case the 
identity (\ref{LP}), thanks to (\ref{ll}) becomes (let me now write $\.G$ 
instead of $D_tG$)
\beq\label{Gll}\nabla_a(\.G)\=\la(J\Phi)_a\=-\la\,\nabla_a\,G\ .\eeq
In addition, in this context, it may be useful to introduce an extension 
of the notion of $\la$-\sy , replacing the scalar function $\la$ with a 
$n\times n$ matrix $\Lambda$ (depending in general on $q,p,\.q,\.p,t$): 
the $\la$-\sy\ condition 
(\ref{ll}) for the DS $\.u=F(u)=J\nabla H$ is replaced by (see \cite{PLA,JN})
\beq\label{LL} [F,\Phi]_a\=-(\Lambda\Phi)_a\eeq
and the above identity (\ref{LP}) must be modified accordingly, giving 
the following
\begin{Proposition}
If a DS admits $X=J\nabla G\cdot\nabla$ as a $\Lambda$-\sy , then $\.G$ obeys  the 
\eq
\beq\label{LA}\nabla_a(\.G)\=(J\,\Lambda\, \Phi)_a\=(J\,\Lambda\,J\,\nabla)_a\,G \ .\eeq
\end{Proposition}
The  two \eq s (\ref{Gll},\ref{LA}) clearly point out an interesting 
property of $\la$ (or $\Lambda$) \sys : they can be viewed as 
``perturbations''  of the ``exact'' \sy . Equations (\ref{Gll},\ref{LA}) 
indeed express the ``deviation'' from the exact conservation  rule $\.G=0$ 
produced by the presence of a nonzero $\la$ (or $\Lambda$).

Some examples will clarify this point. The first one deals with the case 
of a $\la$-\sy\ (i.e. with a scalar function $\la$); it is quite simple and can 
be useful to illustrate the idea. The second one deals with a  
Toda-like Hamiltonian and a $\Lambda$-\sy\ with the introduction of a $\Lambda$ 
matrix. In both cases, the deviation from the exact conservation of $G$ 
will be explicitly evaluated and the ``controlled'' failure of the 
conservation rule clearly described.

\medskip\noindent
{\it Example 3.}
Consider the Hamiltonian in $m=2$ degrees of freedom 
\[H\=\frac 1 2 p_1^2+\frac 1 2 p_2^2+\frac 1 2 q_1p_1^3+\frac 1 2 
q_2^2p_1^2\ .\]
The vector field $X=\pd/\pd q_1$ is a $\la$-\sy\ for the Hamilton \eq s 
of motion $\.u=F(u)=J\nabla H$ 
(which can be easily written), with $\la$ given by the scalar function 
$\la=3p_1^2/2$, namely
\[ [F,\Phi ]\=-\frac 3 2 p_1^2\Phi\q\q , \q\q \Phi=(1,0,0,0)^t \ .\]
Expectedly, the generating function $G=p_1$ is not conserved, indeed one 
has
\[\.G\=-\frac 1 2 G^3\=-\frac 1 3 \la\, G\ .\]
Elementary integration gives $G(t)=G_0\big(1+t\,G_0^2\big)^{-1/2}$ with
$G_0=G(0)$, and
\[\.G\=-\frac 1 2  G_0^3   \big(1+t\,G_0^2\big)^{-3/2}\]
which precisely expresses ``how much'' $G$ is not conserved and indicates 
in particular that $G$ is ``almost conserved" for great values of $t$.

\medskip\noindent
{\it Example 4.}
Consider now the following $2$ degrees of freedom Toda  Hamiltonian
\[ H\=\frac 1 2 p_1^2+\frac 1 2 p_2^2+e^{q_1+q_2}+e^{q_1-q_2}\ .\]
It is easy to write the corresponding Hamilton \eq s of motion and to verify 
that the vector field
\[ X\=\frac{\pd}{\pd q_1}+\frac{\pd}{\pd q_2}\]
with generating function $G=p_1+p_2$ is a $\Lambda$-\sy\ for this system with 
$\Lambda$ given by the $4\times 4$ matrix
\[ \Lambda\=-2\,e^{q_1+q_2} \pmatrix {0 & 0 \cr I_2 & 0} \]
namely 
\[ [F,\Phi]\=-\Lambda\Phi \q\q ,\q\q \Phi=(1,1,0,0)^t\ .\]
In agreement with the above discussion and Proposition 7, one obtains
\[ \.G=\.p_1+\.p_2=-2\,e^{q_1+q_2}\q\q ,\q\q 
\nabla(\.G)=-2\,e^{q_1+q_2}\Phi\ .\]
Introducing the variables
\[w_1=q_1-q_2 \q ,\q w_2=p_1-p_2 \q ,\q w_3=G=p_1+p_2 \q ,\q \z=q_1+q_2\]
the DS becomes
\[ \.w_1=w_2 \q ,\q \.w_2=-2e^{w_1} \q ,\q \.w_3=-2e^\z \q ,\q \.\z=w_3\ .\]
It can  be noted that this DS has not the ``reduced" form as said in 
Proposition 3 and eq. (\ref{wz}), where $\z$ is present only in the r.h.s. of $\.\z$: 
indeed the reduced form (\ref{wz}) 
is granted only if $\Lambda$ is a scalar, $\Lambda=\la I$. Anyway, the 
system is easily solvable; in particular one has 
\[\z\=2\log\Big(\frac {|c_1|}{\cosh (c_1t+c_2)}\Big) \q , \q \.G\=-2e^\z \]
\[ G\=-2c_1\tanh(c_1t+c_2) \q ,\q  |\.G(t)|\le 2 |c_1|\]
where $c_1,c_2$ are arbitrary constants, which shows that $G$ is not a 
conserved quantity, as expected, however -- for any choice of  
$c_1,c_2$ -- both $\.G(t)$ and $G(t)$ are determined and bound quantities.

\section{Conclusion}

Finding Lie point-\sys\ of a dynamical system is in general a quite 
difficult task, and it is strictly connected with the searching for its constants of motion.
In this paper I have discussed and emphasized this close connection, using old and new 
results. I have shown in particular that the knowledge of a \sy\ of the 
DS can allow to directly obtain quantities which are conserved and invariant 
under the \sy . 

The particular case of Hamiltonian DS is specially 
interesting: a classical result ensures that if  a \sy\ of the DS admits a 
generating function $G$, then $G$ is automatically a constant  of motion and a 
\sy-invariant. If instead the Hamiltonian DS admits a \sy\ of ``weaker'' 
type (specifically, a $\la$ or a $\Lambda$-\sy ) then the generating function 
is no longer a conserved quantity, but we have seen that the deviation from the exact 
conservation rule $\.G=0$ is ``controlled'' in a well defined way.

\end{document}